\documentclass[]{spie}  

\usepackage{amsmath,amsfonts,amssymb}
\usepackage{graphicx}
\usepackage[colorlinks=true, allcolors=blue]{hyperref}
\usepackage{booktabs} 
\usepackage[raggedrightboxes]{ragged2e}

\title{Accounting for Clock-Induced Charge Production in the EMCCD Gain Register}

\author[a]{Kevin J. Ludwick}
\affil[a]{University of Alabama in Huntsville, 301 Sparkman Dr, Huntsville, AL 35899, USA}

\authorinfo{Further author information: (Send correspondence to K.J.L.)\\K.J.L.: E-mail: kevin.ludwick@uah.edu, Telephone: 1 256 824 2527}

\begin{document}
\vspace{1.5cm}
\vspace{1.5cm}
\vspace{1.5cm}
\maketitle
\newpage
\begin{abstract}
An electron-multiplying charge-coupled device (EMCCD) is often used for taking images with space telescopes and other devices.  Photons hit the pixels and photo-electrons are created, and these are multiplied via impact ionization as they travel through the gain register from one gain stage to the next.  A high gain means a high multiplication factor, and this is achieved through a high voltage difference across a gain stage.  If the gain is high enough, the chance of clock-induced charge production in the gain register increases.  The probability distribution function governing the gain process in the literature only accounts for charge multiplication if one or more electrons enters the gain register.  I derive from first principles the modified probability distribution that accounts for clock-induced charge production in the gain register.  I also examine some EMCCD data and show through maximum likelihood estimation that the data conform better to the modified distribution versus the usual one in the literature.  The use of the modified distribution would in principle improve the accuracy of signal extraction from a frame.
\end{abstract}


\section{INTRODUCTION}
\label{sec:intro}  

An electron-multiplying charge-coupled device (EMCCD) is useful for low-signal astronomy.  The photon signal from a low-signal observation target that has low contrast must compete with noise sources such as clock-induced charge, dark current, and read noise, where read noise is the most significant source and typically dominates over the signal.  An EMCCD multiplies the photo-electron counts in each pixel via a gain register, which is a series of stages that each pixel's charge proceeds through before read-out.  Each stage accelerates the charges through a voltage difference, and the charges that enter the stage gain more charges due to impact ionization, so more charges leave the stage than entered.  Let $P$ be the probability that a charge turns into two via impact ionization in a gain stage, and assume this probability per charge is a constant (or assume it is the average probability of multiplication by two in a stage).  The amplification, or gain, is typically modeled as follows\cite{Robbins}:
\begin{equation}
g = (1+P)^M,
\end{equation}
where $g$ is the gain (factor of multiplication of the charges that exit the gain register relative to the number that entered) and $M$ is the number of gain stages.

The probability $p^1_M$ that $x$ charges exit a gain register with $M$ stages when a single charge enters the gain register is given in Basden (2003)\cite{Basden}, and it originates from Matsuo (1985)\cite{Matsuo}:
\begin{align}
\label{Basden1}
& p^1_M(x) = (1-P) p^1_{M-1}(x) + P \sum_{k=0}^{x} p^1_{M-1}(x-k) p^1_{M-1}(k), ~~ x,M \geq 1 \nonumber \\
& p^1_M(0) = 0, ~~~ M \geq 1 \nonumber \\
& p^1_0(x) = \delta_{1,x}, ~~ M \geq 1.
\end{align}
The probability that a charge in a stage does not multiply is $(1-P)$, so the only way to get $x$ outgoing charges if no multiplication occurs in the $n$th stage is if $x$ charges come into the $n$th stage from the $(n-1)$st stage, as represented in the first term of the top line of the previous equation.  The terms in the sum represent all the ways in which an extra charge may be added (with probability $P$).  For a given charge in a gain stage, it will either multiply, with probability $P$, or not multiply, with probability $(1-P)$.  Therefore, each possible interaction scenario will consist of a product of some number of factors of $P$ and some number of factors of $(1-P)$.  One can calculate the mean expected value of this distribution and find that 
\begin{equation}
\label{prob1}
\sum_{x=0}^{2^M} x~ p^1_M(x) = (1+P)^M,
\end{equation}
which indeed is the gain $g$.  The sum's upper limit is $2^M$ since that is the biggest number of charges that can possibly result from this model of the gain register.  This probability distribution can be approximated by the exponential distribution if $M$ is large and $P$ is small\cite{Basden}:
\begin{equation}
\label{exp}
p^1_M(x) \approx \frac{e^{-x/g}}{g},
\end{equation}
which has a mean expected value of $g$.  Note that a high value of $g=(1+P)^M$ is still attainable for small $P$ as long as $M$ is large enough.

Eq. (\ref{prob1}) assumes the only way a charge can multiply is by producing one and only one extra charge in a stage.  However, this can be generalized to account for the probability of multiplication by three (production of two charges by an incoming charge).  If one calls that probability $P_2$, then the probability distribution function can be expressed as 
\begin{align}
\label{prob2}
& p^1_M(x) = (1-P-P_2) p^1_{M-1}(x) + P \sum_{k=0}^{x} p^1_{M-1}(x-k) p^1_{M-1}(k) + P_2 \sum_{k=0}^x \sum_{l=0}^k p^1_{M-1}(x-k) p^1_{M-1}(k-l) p^1_{M-1}(l), ~~ x,M \geq 1 \nonumber \\
& p^1_M(0) = 0, ~~~ M \geq 1.  \nonumber \\
& p^1_0(x) = \delta_{1,x}, ~~ M \geq 1.
\end{align}
Physically, $P_2< P$.  The mean expected value of this distribution is
\begin{equation}
g = (1+P+2P_2)^M.
\end{equation}
Similar modifications will enable the inclusion of multiplication by four and five and so on.
However, one can simply absorb all probability terms into another constant $\tilde{P}$ and write 
\begin{equation}
\label{tildeP}
g = (1+\tilde{P})^M.
\end{equation}
Therefore, multiplication by two or more can be written as an effective multiplication by two with a different probability of impact ionization $\tilde{P}$.  Furthermore, I only intend to use a convenient, widely-used approximation for this distribution that depends on $g$ and not on the exact distribution, which would differ some when the inclusion of multiplication by three or more.

As Basden (2003) points out, the probability distribution for two charges entering the gain register is the convolution $p^1_M(x) * p^1_M(x)$, and the distribution for $n$ incoming charges for $M$ stages is\cite{Basden}
\begin{equation}
\label{prob_conv}
p^n_M(c) = p^{n-1}_M(x) * p^1_M(x) = \sum_{x=n-1}^c p^{n-1}_M(x) \times p^{1}_M(c-x), ~~~ c \geq n, M \geq 1,
\end{equation}
where the lower limit of the sum is correct and differs from what is in Basden (2003), which is $x=n$.\footnote{This can be easily checked in the case of, for example, $c=n$, where $p^n_M(c)$ would be 0 if the lower limit of the sum were $x=n$, whereas the probability is expected to $(1-P)^M$.  This is because, for $n$ incoming charges, if $n$ charges leave the gain register, no multiplication in any stage would have occured, and the probability that this situation occurs is $(1-P)$ times itself $M$ times.}  As a result, the expressions in Eq. (A4) in Basden (2003) differ from what is below.

Using Eq. (\ref{exp}) as an approximation for $p^1_M(x)$, one can convolve it with itself a few times and see that a pattern emerges.  $p^n_M(x)$ is found to be
\begin{equation}
\label{Pnex}
p^n_M(x) \approx \frac{(x-(n-2))  e^{-\frac{x}{g}} \left(\prod _{i=2}^{n-1} (i+x)\right)}{(n-1)! g^n}, ~~ x \geq n.
\end{equation}
This can be evaluated to be
\begin{equation}
\label{Pnexact}
p^n_M(x) \approx \frac{g^{-n} e^{-\frac{x}{g}} (-n+x+2) \Gamma (n+x)}{\Gamma (n) \Gamma (x+2)}, ~~ x \geq n.
\end{equation}
For large $x$, keeping only the highest order term in $x$, one can further approximate the distribution to obtain
\begin{equation}
\label{Erlang}
p^n_M(x) \approx \frac{x^{n-1}  e^{-\frac{x}{g}}}{g^n (n-1)!}, ~~ x \geq n, ~x \gg 1,
\end{equation}
This approximate expression is convenient, especially for typical EMCCD usage in which gain is large and thus $x$ is large.  However, it is not normalized as it is, although it is approximately so.  If $x \in \mathbb{R}$ and $x \in (0, \infty)$, then this is the Erlang distribution (the Gamma distribution for integer-valued $n$), and the Erlang distribution is normalized, and its expected mean value is $g$.  However, $x$ is instead specified to be integer-valued ($x \in \mathbb{N}$), and $x \geq 1$.  In that case, the sum of Eq. (\ref{Erlang}) from $x=n$ to infinity is 
\begin{equation}
\label{HurwitzLerch}
\frac{\left(e^{-1/g}\right)^n g^{-n} \Phi \left(e^{-1/g},1-n,n\right)}{(n-1)!},
\end{equation}
where $\Phi$ is the Hurwitz-Lerch transcendent.  This expression is $1$ to machine precision for large-enough values of $n$ and $g$, and it is very close even for smaller values.  For example, for $n=5$ and $g=10$, Eq. (\ref{HurwitzLerch}) gives $0.999898$.

The more general approximate distribution Eq. (\ref{Pnexact}) is not normalized either.  The sum over the range (from $x=n$ to infinity) is
\begin{equation}
\label{Pnexactnorm}
\frac{4^n e^{-\frac{n+1}{g}} g^{-n} \Gamma \left(n+\frac{1}{2}\right) \left(e^{1/g} \, _2\tilde{F}_1\left(1,2 n;n+2;e^{-1/g}\right)+n \, _2\tilde{F}_1\left(2,2 n+1;n+3;e^{-1/g}\right)\right)}{\sqrt{\pi }},
\end{equation}
where $_2\tilde{F}_1(a,b;c;z)$ is the regularized hypergeometric function and equals $_2F_1(a,b;c;z)/\Gamma(c)$.  However, this normalization factor is very close to 1.  For example, for $n=1$ and $g=1000$, this gives 0.9995, and the expression gets closer to 1 the bigger that $g$ is.  Since partial CIC should only be prominent at large gain values, the normalization factor is not so important, although I do employ it when feasible in the proceeding sections.

Brian Sutin (2023) \cite{Sutin} develops a nice matrix formalism for representing the probability distribution for $n=1$ (Eq. (\ref{Basden1})) and $n>1$ (Eq. (\ref{prob_conv})).  If one wants to avoid the use of approximations to the exact probability distributions, his formalism presents a more straightforward way to calculate without approximating Eq. (\ref{prob_conv}).  However, the difference between the approximate forms I show here and the exact form is only significant for very large $x$, far above the mean.  These approximate forms are easier to work with and are sufficient for my purposes, especially since partial CIC has the biggest effect at relatively small values of $x$, as will be shown. 

In summary, $p^n_M(x)$ accounts for charges that entered the gain register and charges produced as a result of those incoming charges via impact ionization.  However, the distribution does not account for charge multiplication as a result of charges that did not enter the gain register, i.e., clock-induced charges (CIC) created in the gain register and their progeny via impact ionization.  The production of such charges has been studied and quantified for an EMCCD by Bush {\it et al} \cite{Bush}.  During 
clocking in a gain stage, holes can be created on the silicon chip.  These holes will be accelerated due to the voltage difference in the gain stage, and they can collide with silicon atoms to create an electron.  This electron would then pass on to the remaining gain stages and multiply like all the other non-CIC electrons.  High gain means a high voltage difference, so the effect of CIC created in the gain register would be most noticeable in high-gain frames of an EMCCD.  CIC produced before the gain register get multiplied as usual with all other charges that enter the gain register.  Since the gain-register CIC travels through fewer gain stages, this phenomenon can be called ``partial CIC".

In this work, I will derive the probability distribution that includes clock-induced charges created in the gain register.  I use maximum likelihood estimation (MLE) with real EMCCD data and show that the distribution which accounts for partial CIC results in a higher likelihood than the distribution from the literature that does not.  Others have used MLE for estimating the gain applied to a frame, as opposed to the gain commanded \cite{Ryan}, albeit without the consideration of partial CIC.  Others have modeled CIC created in the gain register \cite{}, but I contend that my approach here is more accurate as it is from first principles as much as practicable.  For high-gain data, the fitted parameter governing the prevalence of partial CIC is much higher than it is for low-gain data, as expected.  I then discuss the noise (unwanted electrons) due to partial CIC.  MLE is one way to estimate the actual gain that was applied to a frame versus the gain commanded to the EMCCD.  Knowing the actual gain is useful for calibrating an EMCCD.  Accounting for noise due to partial CIC, along with an accurate estimate of the gain, is useful for accurately extracting the photon signal from an observation target. 


\section{PROBABILITY DISTRIBUTION FOR PARTIAL CIC}

In the previous section, $p^2_M(x)$ was obtained by convolving $p^1_M(x)$ with itself, which assumed that the presence of a second incoming charge to the gain register and all its progeny were independent of the first charge and its progeny.  I will also assume independence of partial CIC from the incoming particles and their progeny.  First, I will consider the probability distribution for partial CIC alone (i.e., no incoming charges into the gain register), and later I will combine partial CIC with the probability distribution from the previous section $p^n_M(x)$ via convolution.  

Assume a gain register with a gain of $g$ with $M$ stages.  Let the probability that a CIC is created in gain stage 1 be $Q$.  Physically, the chance of the production of a CIC ($Q$) is lower than the probability of production via impact ionization ($P$).  For now, I will ignore the prospect of CIC creation in all other gain stages.  In that case, the probability that $x\geq 1$ charges exit the gain register after a CIC is created in gain stage 1 is $Q p^1_{M}(x)$, or $Q$ times the probability distribution for 1 charge entering a set of $M$ stages.  In this case, once the charge is created, it follows the distribution for impact ionization, which is possible in all of the $M$ stages.  The probability of producing more than one CIC in a single gain stage is assumed to be negligible.  In that case, the probability that 0 charges exit the gain register is $(1-Q)$\footnote{For the inclusion of the case where there is a non-zero probability of more than one CIC produced for a single gain stage, one could most likely argue for an effective probability $\tilde{Q}$ in a manner similar to what was used to absorb the probability terms into $\tilde{P}$ in Eq. (\ref{tildeP}).}.  One can write the probability for any $x$ in this case as
\begin{equation}
Q p^1_{M}(x) + (1-Q) \delta_{0,x},
\end{equation}
where $\delta_{0,x}$ is the Kronecker delta.   
Of course, other CIC can be produced in other gain stages.  To include the cases of CIC created in other gain stages and their progeny, one must convolve the expression with the case for gain stage 2 with $M-1$ stages, gain stage 3 with $M-2$ stages, and so on.  The result is
\begin{equation}
\label{multi_conv}
q^0_{M}(x) =  [Q p^1_{M}(x) + (1-Q) \delta_{0,x}] * [Q p^1_{M-1}(x) + (1-Q) \delta_{0,x}] * [Q p^1_{M-2}(x) + (1-Q) \delta_{0,x}] *  \dots * [Q p^1_{1}(x) +(1-Q)\delta_{0,x}],
\end{equation}
where I use $q$ to denote the probability distribution for CIC.  The superscript denotes that there are 0 incoming particles to the gain register, and the subscript indicates $M$ gain stages in the register.  
To expand this, first notice that
\begin{equation}
\delta_{0,x} * \delta_{0,x} = \sum_{i=0} \delta_{0,i} \delta_{0,x-i} = \delta_{0,x},
\end{equation}
and also
\begin{equation}
p^1_{r}(x) * \delta_{0,x} = \sum_{i=0}^x p^1_r(x) \delta_{0,x-i} = p^1_r(x).
\end{equation}
Using these two facts, one can expand Eq. (\ref{multi_conv}).  I first show an example, where $M=3$ (suppressing the notation for $x$ dependence):
\begin{align}
\label{3ex}
q^0_3 = &Q^3 p^1_3 * p^1_2 * p^1_1 + Q^2(1-Q) p^1_3 * p^1_2 + Q^2(1-Q) p^1_3 * p^1_1 + Q(1-Q)^2 p^1_3  + Q^2(1-Q) p^1_2 * p^1_1 \\ \nonumber & + Q(1-Q)^2 p^1_2 + Q(1-Q)^2 p^1_1 + (1-Q)^3 \delta_{0,x}.
\end{align}
One can see that the result is the sum of the convolution of $p^1_r$ with all possible convolution combinations involving $r-1$ down through $1$ for all sizes of selections ranging from $i=r-1$ down to $i=0$, where $r$ ranges from 1 to $M$. For $r=M=3$, this means the first four terms:  {321} ($i=2$), {32, 31} ($i=1$), {3} ($i=0$).  For $r=M-1=2$, this means the fifth and sixth terms:  {21} ($i=1$), {2} ($i=0$).  For $r=M-2=1$, this means the seventh term:  {1} ($i=0$).  For each term, the number of factors of $Q$ is $i+1$, and the number of factors of $(1-Q)$ is $M-(i+1)$.  Finally, there is the last term for the convolution of all the delta functions, with $M$ factors of $(Q-1)$. 

Using Eq. (\ref{Basden1}) to compute the convolution of factors of $p^1_r$ with different subscript values is possible but certainly unwieldy and unnecessary.  A quantification of partial CIC from the data is possible without such a degree of precision and complexity.  Instead, I use the approximate form from Eq. (\ref{Pnexact}) for the $p^1_r$ factors.  This approximate form is not really valid for small subscript values.  
However, the terms 
with high subscript values contribute the most to partial CIC, as partial CIC is more likely to be produced from a high number of gain stages than from a low number (i.e., good chance of large progeny), so I use this approximation for all $p^1_r$ factors.  

Notice that $p^n_M$ from Eq. (\ref{Pnexact}) uses $g$ for $M$ gain stages.  Since $g=(1+P)^M$, I can substitute $g^{r/M}$ for $g$ in Eq. (\ref{Pnexact}) to represent the distribution for $r<M$ gain stages.  I could in theory compute the convolution for every term that appears in Eq. (\ref{3ex}) assuming the form from Eq. (\ref{Pnexact}), but again, that level of precision and complexity is unnecessary; I aim for a manageable formula for partial CIC.  Instead, I will approximate each term involving a convolution as $p^n_{\tilde{r}}$, where $n$ is the number of factors involved in the convolution for the term, and $\tilde{r}$ is the approximate, effective number of gain stages for the approximate probability distribution resulting from the convolution.  The convolution of $n$ factors of $p^1_r$ will result in a probability distribution for $n$ incoming charges, as shown in the previous section.  The derivation of $\tilde{r}$ is below.

I approximate an effective number of gain stages $\tilde{r}$ for the convolution as follows.  I consider all terms whose biggest subscript value is the same, say, $r$.  For a general value $r$, the number of terms with $p^1_r$ with $i$ factors convolved with it  (what I called ``size selection" above) is equal to the number of ways $i$ unique elements of the set $s=(1, 2, 3, ..., r-1)$ can be chosen.  For example, for $r=3$ and a size selection of $i=1$, there are two terms, the second and third terms from Eq. (\ref{3ex}), and the average value of the subscript values is the average number of gain stages for each term ($5/2$ and $2$ respectively).  For general $r$, the number of terms will be large and unwieldy, if analytic.  To keep things simple, I average the average subscript value over all the terms.  The average number of gain stages for this size selection $i$ is $r$ plus the average value of the sum of the $i$ elements over all the terms under consideration, all divided by the number of factors in each term.  For the example above ($r=3$ and $i=1$), I use $p^n_{\tilde{r}} = p^2_{9/4}$ to approximate each term (2 for the superscript since there are two factors in each term with a superscript of 1 and $\tilde{r} = [3+ (2+1)/2]/2 =9/4$, where $(2+1)/2$ is the average subscript value found in the set $s$ and the subscript in common, 3, is averaged with it over the number of factors in these terms, which is 2).  Since the selection from $s$ is unique for each term, the sum of $i$ elements ranges from $l=1+2+3+...+i$ to $h=(r-1)+(r-2)+(r-3)+...+(r-i)$, with every integer in between covered by a term exactly once.   
Therefore, the average number of gain stages is 
\begin{equation}
\tilde{r} = \frac{r + \frac{l+h}{2}}{i+1} =  \frac{r + \frac{\sum_{j=1}^i j + \sum_{j=1}^i (r-j)}{2}}{i+1} = \frac{r + \frac{\sum_{j=1}^i r}{2}}{i+1} = \frac{r+\frac{ir}{2}}{i+1}.
\end{equation}
For a given value of $r$, there are ${r-1 \choose i}$ terms coming from selections from $s$.  The result of summing up contributions from all values of $i$ for a given $r$ is
\begin{equation}
\sum^{r-1}_{i=0} {r-1 \choose i} p^{i+1}_{\tilde{r}} (1-Q)^{M-(i+1)} Q^{i+1}.
\end{equation}
This sum is not closed and analytic, and this is because of the $i$-dependence of $\tilde{r}$.  One can try taking a representative value for $\tilde{r}$ to get the terms in the sum that contribute most.  Of all possible values of $i$, the one that corresponds to the highest number of ``incoming particles", the superscript, would contribute the most  to partial CIC's progeny.  This would be the highest value of $i$, which is $r-1$.  Substituting $r-1$ for $i$, $\tilde{r}$ becomes\footnote{The mean value of $\tilde{r}$ is $\sum_{i=0}^{r-1} \frac{r+\frac{ir}{2}}{i+1}/(r-1) = \frac{r \left(H_r+r\right)}{2 (r-1)}$, where $H_r$ is the $r$th harmonic number.  For large values of $r$, this agrees well with $\frac{r+1}{2}$.}
\begin{equation}
\frac{r+1}{2},
\end{equation}
which is what I use to obtain a closed, analytic form for the sum:
\begin{align}
\label{H}
&H(r,x) \equiv \sum^{r-1}_{i=0} {r-1 \choose i} p^{i+1}_{(r+1)/2} (1-Q)^{M-(i+1)} Q^{i+1} &\\ \nonumber
& = Q (1-Q)^{M-2} g^{-\frac{r+1}{M}} e^{-x g^{-\frac{r+1}{2 M}}} \bigg(Q (1-r) \, _2F_1\big(2-r,x+2;2;\frac{g^{-\frac{r+1}{2 M}} Q}{Q-1}\big)& \\ \nonumber 
&+(1-Q) g^{\frac{r+1}{2 M}} \, _2F_1\big(1-r,x+1;1;\frac{g^{-\frac{r+1}{2 M}} Q}{Q-1}\big)\bigg).&
\end{align}
To compute the sum, I used Eq. (\ref{Pnexact}) for $p^{i+1}_{(r+1)/2}$, using $g^{(r+1)/(2M)}$ as the corresponding effective gain.  In order to get this closed, analytic result, I used Eq. (\ref{Pnexact}) without dividing by the normalization factor from Eq. (\ref{Pnexactnorm}), which I argued was approximately 1 for high gain.  

$H(r,x)$ covers all the terms with biggest subscript $r$, and again, $r$ ranges from $1$ to $M$.  Putting it all together, the approximate probability distribution for no incoming particles into the gain register, accounting for partial CIC, is
\begin{equation}
\label{q}
q^0_M(x) = \sum^M_{r=1} H(r,x) + (1-Q)^M \delta_{0,x},
\end{equation}
which can also be written as
\begin{equation}
q^0_M(x) =
	\begin{cases} 
		& \sum^M_{r=1} H(r,x), ~~ x>0 \\
		& (1-Q)^M, ~~ x=0.
\end{cases}
\end{equation}
The sum in Eq. (\ref{q}) cannot be computed analytically without making some approximation.  It is possible to do a Taylor expansion assuming $Q\ll1$ and additionally approximate the sum as an integral to get an analytic approximation for the sum.  One can also compute an approximate normalization factor by integrating the result over $x \in (1,\infty)$.  
However, the result to first order in $Q$ deviates significantly enough from the brute-force numerical (truncated) sum that we employ the numerical sum in our analysis code for performing MLE, both for the sum in Eq. (\ref{q}) and for the corresponding normalization factor.  I have verified in my analysis (discussed later) that the normalization factor is very close to 1 as expected, but because of the approximations made before Eq. (\ref{q}), it is best to compute the normalization factor and divide by it for the sake of accuracy, especially for MLE, for which having a normalized probability distribution is important.

\begin{figure}[h!]
\begin{center}
\includegraphics[scale=0.6]{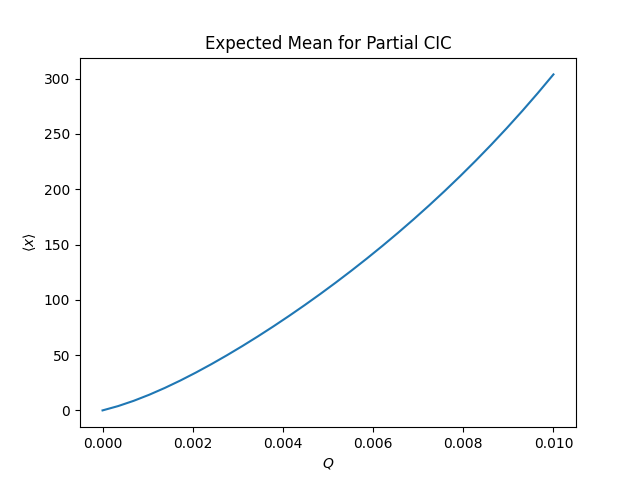}
\end{center}
\caption{ The expected mean for the partial CIC distribution as a function of $Q$, the probability of CIC production in a given gain stage.  The parameters used are $M=604$ and $g=5000$.  Realistically, $Q<P$, where $g=(1+P)^M$, so the domain shown in the plot is limited accordingly.}
\label{mean_plot}
\end{figure}

\begin{figure}[h!]
\begin{center}
\includegraphics[scale=0.6]{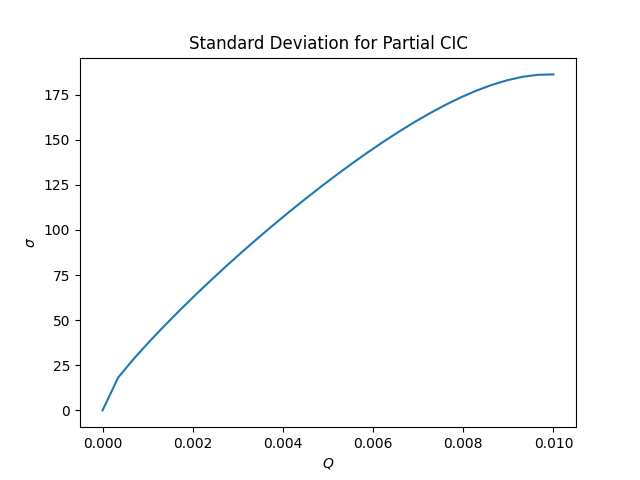}
\end{center}
\caption{ The standard deviation for the partial CIC distribution as a function of $Q$, the probability of CIC production in a given gain stage.  The parameters used are $M=604$ and $g=5000$.  Realistically, $Q<P$, where $g=(1+P)^M$, so the domain shown in the plot is limited accordingly.}
\label{std_plot}
\end{figure}

Fig. (\ref{mean_plot}) shows the expected mean, $\langle x \rangle$, for the normalized partial CIC distribution ($q^0_M(x)$) for a typical high-gain scenario, and Fig. (\ref{std_plot}) shows the standard deviation, $\sigma = \sqrt{\langle x^2 \rangle - \langle x \rangle^2}$.

Eq. (\ref{q}) gives the probability distribution for 0 charges entering the gain register, so the charges that exit are solely due to partial CIC.  If charges entered the gain register, they would multiply in the gain stages, independent of partial CIC and its progeny.  Therefore, the distribution for $n$ incoming particles is (using $|$ to separate the variables from the fixed parameters)
\begin{equation}
q^n_M(x|Q,g, n) = q^0_M(x | Q,g) * p^n_M(x| g,n).
\end{equation}
In my analysis, I use Eq. (\ref{Pnexact}) divided by the normalization factor from Eq. (\ref{Pnexactnorm}) for $p^n_M(x)$, and I compute the convolution using a fast Fourier transform (FFT):  I compute $q^0_M(x|Q,g)$ and $p^n_M(x|g,n)$ analytically using the equations given in this work over an array of values at a sufficiently high sampling which contains the relevant range of $x$ corresponding to the data to perform MLE on, and then I perform a FFT on each array, multiply them together, and perform the inverse FFT on the result.  

\section{INCORPORATING PARTIAL CIC INTO READOUT STATISTICS}

Now that the statistics of partial CIC has been developed, it can be incorporated into the statistics of what is actually read out digitally from an EMCCD detector.  

The conversion of a photon to a photo-electron as it interacts with a detector pixel (before the application of EM gain) is a Poisson process, and so is the generation of dark current
and CIC.  If the mean expected number of electrons per pixel 
(photo-electrons as well as noise electrons)
is $\lambda$, the number of electrons $n$ actually produced per pixel before EM gain is given by the Poisson distribution function,
\begin{equation}
\label{Poisson}
P_p (n | \lambda) = \frac{\lambda^n e^{-\lambda}}{n!}.
\end{equation}
The electron counts per pixel are then amplified as
the detector frame of pixels goes through the gain register.  Therefore, $n$ in Eq. (\ref{Pnexact}) is a Poisson variate.  The relevant photo-electron signal in the pixel resulting from the observation of  some luminous source for time $t_{fr}$ is $s$.  The expected value of $s$ is $\langle s \rangle = \phi \eta t_{fr}$, where $\phi$ is the photon flux (photons/s), $t_{fr}$ is the exposure time for the frame, and $\eta$ is the pixel's
quantum efficiency (electrons/photon).    Along with this photo-electron signal, dark current $i_d$ and clock-induced charge $CIC$ contribute to the number of electrons in this pixel.   The Poisson variate $n$ can be written as the sum of the aforementioned Poisson variates:
\begin{equation}
n = s + i_d t_{fr} + CIC.
\end{equation}
The frames I study in this work are dark frames (taken in the absence of a luminous source), and so $n$ is mainly due to the second two terms.  I avoid "bright" frames (in the presence of a luminous source) to avoid having uneven distribution of light on the pixels; that way, it is more likely that a single set of parameters can describe the statistics of a histogram of the frame for all the pixels, which is useful for performing MLE with the histogram.

For a given pixel in a given frame, the probability of getting $x$ electrons after the application of EM gain given a mean number of pre-gain counts $\lambda$ is given by the composition
\begin{equation}
\label{P_Pg}
P_{Pg} (x | Q, g, \lambda) =  \sum_{i=0}^{\infty} q^n_M(x | Q, g, i) P_p(i | \lambda). 
\end{equation}
If computing the sum numerically, the sum is usually approximated well by truncating to about five standard deviations above the mean (i.e., summing from $i=0$ to $i=5 \sqrt{\lambda}$).  
If no partial CIC is accounted for, this PDF would instead be
\begin{equation}
\label{P_Pg_no}
\textbf{no partial CIC}:  ~~ P_{Pg} (x | g, \lambda) =  \sum_{i=0}^{\infty} p^n_M(x | g, i) P_p(i | \lambda). 
\end{equation}

After the charges have passed through the gain register, they are converted to a digital output, and voltage bias is added to each pixel's value.  While the camera data is read out to a computer that collects the detector data, read noise contributes to the pixel counts as well.  We can model the readout for a single pixel as a random variate $q$:
\begin{equation}
\label{readout}
q = b + x + r,
\end{equation}
where $b$ is the voltage bias that was applied, $r$ is the contribution from read noise and is a variate of the normal distribution,  and $x$ is a variate of the distribution from Eq. (\ref{P_Pg}).  

The average bias can be determined by taking detector frames at very short exposure times.  For these frames, the median of the pixel values is dominated by the bias in units of digital counts (DN).  This amount then can be subtracted from a frame of interest.  Also, while the pixel charge values get digitized, an effective factor is applied to the number of charges to convert from electrons to DN.  The factor to convert back to electrons, called k gain, can be calibrated with a photon transfer curve.  Therefore, after one subtracts the bias from a frame of interest and multiplies by the k gain (electrons/DN), the frame has counts that follow a distribution which is the convolution of the normal distribution $P_{rn}$ (for read noise) \cite{Plakhotnik} with the distribution from Eq. (\ref{P_Pg}):
\begin{equation}
\label{P_rPg}
P_{rPg} (x | \mu, \sigma_{rn}, Q, g, \lambda)  = P_{rn} (x | \mu, \sigma_{rn}) * P_{Pg}(x | Q, g, \lambda),
\end{equation}
where 
\begin{equation}
\label{P_rn}
P_{rn}(x | \mu, \sigma_{rn}) = \frac{1}{\sqrt{2 \pi} \sigma_{rn}} \mathrm{Exp}\left[ \frac{-(x-\mu)^2}{2 \sigma_{rn}^2} \right], 
\end{equation}
and $\mu$ is the mean and $\sigma_{rn}$ is the standard deviation (what is called the read noise).  
In my analysis, I use an FFT to compute this convolution on an well-sampled array, and then I interpolate a spline over the array in order to have a function to use in fitting parameters in the MLE.

\begin{figure}[ht]
\label{diagram}
\begin{center}
\includegraphics[scale=0.6]{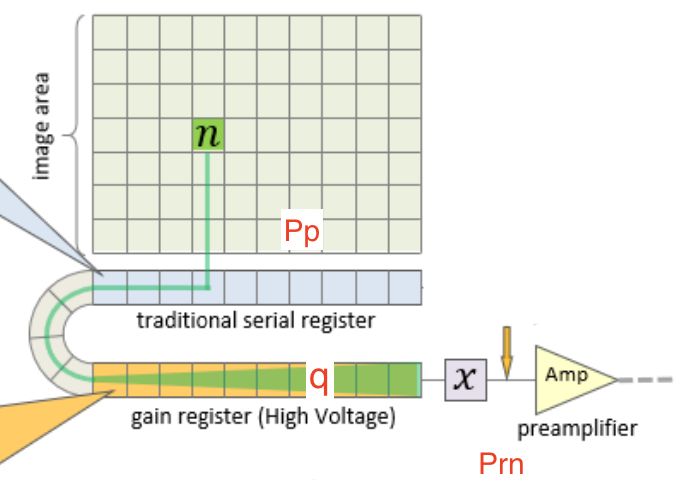}
\end{center}
\caption{ Photons are converted to photo-electrons in the image area, and the photo-electrons follow the Poisson distribution ($P_p$).  The diagram shows $n$ charges in the image area clocked down, row by row, until they reach the serial register.  Then the charges move through the stages of the serial register and then the stages of the gain register, and the statistics here are according to $q^n_M(x|Q,g,n)$, where $n$ is a Poisson variate.  Finally, the charges are read out through a preamplifier, and read noise according to the normal distribution $P_{rn}$ is added.
Image credit:  Edited from Nemati, B., Space Telescopes and Instrumentation 2020: Optical, Infrared, and Millimeter Wave, 11443, 884 – 895, International Society for Optics and Photonics, SPIE (2020).}
\end{figure}

Fig. (\ref{diagram}) summarizes the probability distributions at play and the flow of charges during readout, and Fig. (\ref{P_rPg_plot}) plots $P_{rPg}$ for a few choices of parameters.

\section{PERFORMING MAXIMUM LIKELIHOOD ESTIMATION}

For a given frame, we create a histogram with $N$ bins over all the pixels, after subtracting the bias and converting from DN to electrons.  For bin $k$, let the frequency of that bin be $f_k$.  The likelihood is the product of the probability distribution evaluated at the $f_k$ data points of bin value $k$ over all $N$ bins, and it is a function of the distribution's parameters. The log-likelihood of the probability distribution of a frame, $P_{rPg}$, is
\begin{equation}
\mathcal{L} = \sum_{k=1}^N f_k \ln(P_{rPg}( \theta | k)),
\end{equation}
where $\theta$ represents the free parameters.  In this case, $\theta = (\mu, \sigma_{rn}, Q, g, \lambda)$, and $\theta$ is listed first in the dependencies of $P_{rPg}$ since $P_{rPg}$ is considered a function of those parameters for a fixed $k$ from the histogram data.  For a given frame, I run an optimization algorithm to determine the parameter choices that maximize the log-likelihood, $\hat{\theta} = (\hat{\mu}, \hat{\sigma}_{rn}, \hat{Q}, \hat{g}, \hat{\lambda})$.  For some cases, I only vary a subset of parameters and fix the others, and I will enumerate such cases when I present the results.

For two PDFs with the same number of fitted parameters, the one with the higher log-likelihood (closer to 0 from below) should be the preferred PDF.  When comparing the log-likelihood for two PDFs with different numbers of free parameters, I use the likelihood ratio test, where the $\chi^2$ value is 
\begin{equation}
\label{LRT}
\chi_{21}^2 = -2 (\mathcal{L}_2 - \mathcal{L}_1),
\end{equation}
where $\mathcal{L}_1$ is the log-likelihood for the PDF with $a_1$ fitted parameters and $\mathcal{L}_2$ is the log-likelihood for the PDF with $a_2<a_1$ fitted parameters.  If $\chi_{21}^2$ value is larger than the $\chi^2$ percentile corresponding to a $p$-value of $0.001$ with $a_1-a_2$ degrees of freedom, then the PDF with $a_1$ fitted parameters fits the data better with statistical significance.  

\begin{figure} [ht]
      \begin{center}
      \includegraphics[height=8cm]{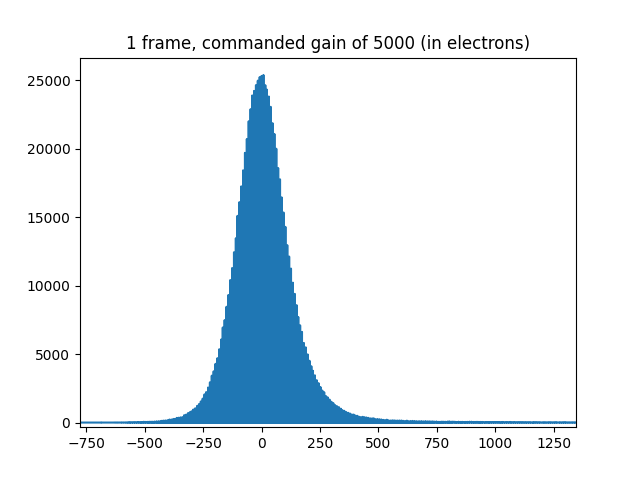}
      \end{center}
      \caption[]
      { \label{data}
      Histogram of a frame taken at a commanded EM gain of 5000.  The tail just to the right of the peak is fatter than expected from the PDF without partial CIC.}
      \end{figure}

The data I examine comes from the T-e2v 301 EMCCD.  Read noise can vary from frame to frame, so to avoid fitting over many frames with many different Gaussian distributions contributing to the read noise, I analyzed a single frame with 1024x1024 pixels, taken with a high commanded EM gain of 5000.  I perform MLE assuming partial CIC (Eq. (\ref{P_rPg})) and also assuming no partial CIC (Eq. (\ref{P_rPg}) using Eq. (\ref{P_Pg_no})).  The assumption of partial CIC yielded better results than without partial CIC.  The shape of the distribution the data follow in Fig. (\ref{data}) resembles the type of shape that can be produced by a sum of Gaussians.  Since the read-out noise can differ even from row to row, the read noise over a whole frame in theory could be due to a sum of Gaussians.  To test whether this shape is due to partial CIC or instead due to multiple Gaussians, I also analyzed a single row of 1024 pixels.  

\begin{figure} [ht]
      \begin{center}
      \includegraphics[height=8cm]{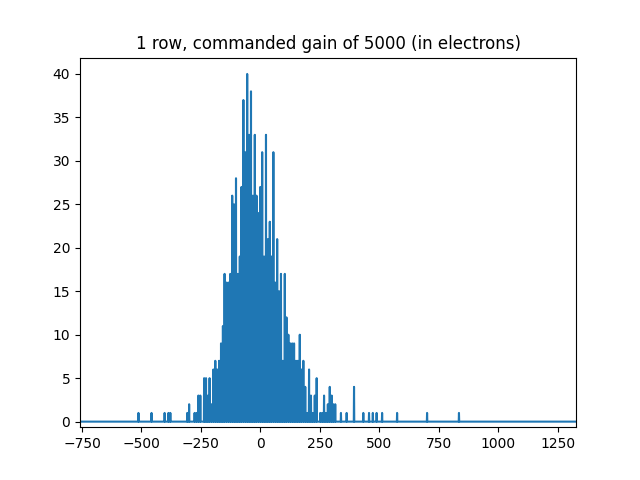}
      \end{center}
      \caption[]
      { \label{data2}
      Histogram of a row of 1024 pixels taken at a commanded EM gain of 5000.  The tail just to the right of the peak is fatter than expected from the PDF without partial CIC.}
      \end{figure}

\begin{figure} [ht]
      \begin{center}
      \includegraphics[height=8cm]{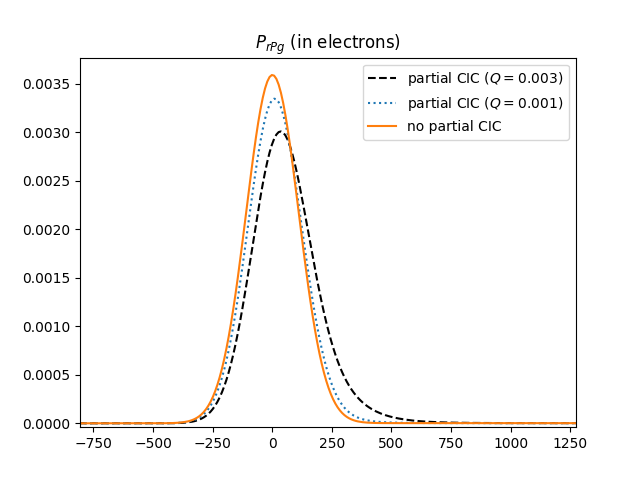}
      \end{center}
      \caption[]
      { \label{P_rPg_plot}
      PDF from Eq. (\ref{P_rPg}), shown with $Q=0$ (no partial CIC), $Q=0.001$, and $Q=0.003$. }   
      \end{figure}

The results are summarized in Tables (\ref{result_table}) and (\ref{result_table2}), and the inclusion of partial CIC was favored overall.  For example, the likelihood ratio test is computed for Fit 2 and the corresponding Fit 4 with the same fitted parameters and the addition of $Q$ for partial CIC.  The likelihood ratio is computed using Eq. (\ref{LRT}), and $\chi_{24}^2 = 14036$.  We compare this to the $\chi^2$ percentile value for $p=0.001$ for 1 degree of freedom, which is 10.83.  Since $\chi_{24}^2 > 10.83$, the fit that includes partial CIC is statistically favored.  Also, note that even though Fit 3 has fewer fitted parameters than Fit 2 has, the log-likelihood of Fit 3 is better than that of Fit 2!  For the single-row fits, Fits 5 and 7 are almost identical; both resulted in $Q=0$.  $\chi^2_{57}$ is very close to 0, so Fit 5 without partial CIC is statistically favored over Fit 7.  However, the inclusion of partial CIC is favored when comparing Fits 6 and 8, when $\sigma_{rn}$ and $\mu$ are allowed to vary.  Overall, the statistics were not very robust for a single row, and this is visually apparent in Fig. (\ref{data2}).  The peak seems to be centered on a negative value further from 0 compared to the apparent peak in the data for a full frame.  The bigger $Q$ is, the more the peak in the distribution shifts to the right because the probability of production of more charges is enhanced due to partial CIC (see Fig. (\ref{P_rPg_plot})).      With an apparent peak so negative, Fit 7 landed on $Q=0$ to avoid a peak shift to the right.  Allowing $\mu$ to vary resulted in a the statistical preference for the inclusion of partial CIC (Fit 8 over Fit 6).  

\begin{table}
  \begin{center}
    \caption{The results of performing MLE on data.  All data is for 1 frame of 1024x1024 pixels.  When read noise is not fitted for, we set $\mu=0e^-$ and $\sigma_{rn}=110e^-$, values from a previous calibration.}
  \label{result_table}
     \begin{tabular}{l | p{3.9in} | p{1in}}
      \toprule 
      {Free Parameters}   & \Centering{Without Partial CIC, Full Frame} & \Centering{Log-Likelihood} \\
      \hline
      \midrule 
       \textbf{1. $\lambda$, $g$} &  \Centering{ $\hat{\lambda}=0.027651e^-, ~ \hat{g}=4971.5$ } &  \Centering{$-6.6766 \times 10^6$}\\
      \textbf{2. $\lambda$, $g$, $\mu$, $\sigma_{rn}$} &  $\hat{\lambda}=0.019089e^-, ~\hat{g}=4861.1, ~\hat{\mu}=5.3154e^-, ~\hat{\sigma}_{rn}=115.32e^-$ &   \Centering{$-6.6665 \times 10^6$} \\
      \hline
	{Free Parameters} &  \Centering{With Partial CIC, Full Frame} &  \Centering{Log-Likelihood} \\
	\hline
	\midrule
	 \textbf{3. $\lambda$, $g$, $Q$} &   $\hat{\lambda}=0.024129e^-,~\hat{g}=5019.2, ~\hat{Q}=1.0588\times 10^{-3}$ &  \Centering{$-6.6624 \times 10^6$} \\ 
       \textbf{4. $\lambda$, $g$, $Q$, $\mu$, $\sigma_{rn}$} & $\hat{\lambda}=0.012357 e^-, ~\hat{g}=5054.9, ~\hat{Q}=1.0529 \times 10^{-3}, ~\hat{\mu}= 3.0188e^-, ~\hat{\sigma}_{rn}=113.02e^-$ &  \Centering{$-6.6595 \times 10^6$}
    \end{tabular}
  \end{center}
\end{table}

\begin{table}
  \begin{center}
    \caption{The results of performing MLE on data.  All data is for 1 row of 1024 pixels to eliminate the potential effect of multiple read noise distributions coming from different rows.  The statistics are less robust than they are for one full frame, but the results still favor the inclusion of partial CIC.  When  read noise is not fitted for, we set $\mu=0e^-$ and $\sigma_{rn}=110e^-$, values from a previous calibration.}
  \label{result_table2}
     \begin{tabular}{l | p{3.9in} | p{1in}}
      \toprule 
      {Free Parameters}   & \Centering{Without Partial CIC, One Row} & \Centering{Log-Likelihood} \\
      \hline
      \midrule 
       \textbf{5. $\lambda$, $g$} &  \Centering{$\hat{\lambda}=0.016783e^-, ~ \hat{g}=5000.0$ } &  \Centering{$-6444.7$}\\
      \textbf{6. $\lambda$, $g$, $\mu$, $\sigma_{rn}$} &  $\hat{\lambda}=0.016520e^-, ~\hat{g}=5000.0, ~\hat{\mu}= -15.527e^-, ~\hat{\sigma}_{rn}=116.15e^-$  &   \Centering{$-6431.7$} \\
      \hline
	{Free Parameters} &  \Centering{With Partial CIC, One Row} &  \Centering{Log-Likelihood} \\
	\hline
	\midrule
	 \textbf{7. $\lambda$, $g$, $Q$} &   $\hat{\lambda}=0.016782e^-,~\hat{g}=5000.0, ~\hat{Q}=0.0000$ &  \Centering{$-6444.7$} \\ 
       \textbf{8. $\lambda$, $g$, $Q$, $\mu$, $\sigma_{rn}$} & $\hat{\lambda}=0.0083818e^-, ~\hat{g}=4999.9, ~\hat{Q}=2.4730 \times 10^{-3}, ~\hat{\mu}=-55.000e^-, ~\hat{\sigma}_{rn}=96.841e^-$ &  \Centering{$-6406.9$}
    \end{tabular}
  \end{center}
\end{table}

\begin{table}
  \begin{center}
    \caption{The results of likelihood ratio tests comparing fits as numbered in Tables (\ref{result_table}) and (\ref{result_table2}).  The ratios are computed for a fit and the corresponding fit with the same fitted parameters with the addition of $Q$ for partial CIC.  We compare to the $\chi^2$ percentile value for $p=0.001$ for 1 degree of freedom, which is 10.83.  When read noise is not fitted for, we set $\mu=0e^-$ and $\sigma_{rn}=110e^-$, values from a previous calibration.}
  \label{chi_table}
     \begin{tabular}{l | p{2.4in} }
      \toprule 
      {$\chi_{ab}^2$ from likelihood ratio test}   & \Centering{Inclusion of Partial CIC Statistically Preferred} \\
      \hline
      \midrule 
       $\chi_{13}^2=28473>10.83$  &  \Centering{Yes} \\
      $\chi_{24}^2=14036>10.83$ &  \Centering{Yes} \\
      $\chi_{57}^2= 2.5622\times 10^{-6}<10.83$ &  \Centering{No} \\
      $\chi_{68}^2=49.580>10.83$ &  \Centering{Yes} \\
    \end{tabular}
  \end{center}
\end{table}

I also performed the same analysis on a set of dark frames taken with a low commanded EM gain, 10.  The partial CIC PDF is in theory perhaps not as accurate as it is in the high-gain case.  It was argued that the terms in $H(r,x)$ in Eq. (\ref{H}) were dominated by the high-$r$ terms, where electron multiplication has a larger effect for a larger number of gain stages, but those high-$r$ terms are not as dominant if the gain is low.  No partial CIC instance fit the data better than an instance without partial CIC in a statistically significant way, but the optimization algorithm always optimized $Q$ to be very small, on the order of $1 \times 10^{-8}$ or less.  This was a good check that the PDF for partial CIC worked reasonably well since $Q$, the probability for the creation of a clock-induced charge in a gain stage, was expected to be low.


\section{Conclusion}

In this paper, I examine the PDF governing the multiplication of charge in the EMCCD gain register and express a general form that is more accurate for low gain values than is typically used in modeling.  I have also derived the contribution to an EMCCD frame coming from clock-induced charge production in the gain register (partial CIC), in a way that is more rigorous than previous attempts.  I then incorporated this into the overall PDF for the output of an EMCCD and used MLE to fit the effect of partial CIC.  I showed that the histogram of the data from the frame matches better visually with the PDF which includes partial CIC.  I also found that the inclusion of partial CIC results in a fit that is statistically favored over its exclusion, demonstrating the reliability of the derivation for partial CIC.  One may argue that the distribution of the data may be affected by different read noise contributions from each row that is read out so that the underlying distribution is partly due to a sum of Gaussians.  However, I showed that partial CIC was preferred even in the case of just one row of a frame, for which read noise should follow a single Gaussian distribution.  For large gain, the voltage across each gain stage is high, and thus the probability $Q$ for CIC in the gain register is higher compared to the case of low gain.  For low-gain frames, I found that $Q$ found via MLE is very low, as expected.  

The clocking scheme in the gain register can be adjusted to minimize partial CIC by minimizing voltage differences and increasing the time duration per stage.  However, one will likely not be able to completely eliminate this effect.  I provide the algorithm for quantifying partial CIC using MLE, and thus it can be used to calibrate in a precise way the commanded gain with the actual gain applied to a frame.  Another application is the improvement of threshold selection in photon counting \cite{Bijan, Hu, Lantz}.  Also, accounting for the effect of partial CIC on the mean and standard deviation of a frame leads to a more accurate prediction of the signal-to-noise ratio (SNR) \cite{Ludwick}, and the planning of observations to achieve a target SNR could be made more accurate with the inclusion of this effect.  For convenience, the code includes the computation for the mean and standard deviation, as illustrated in Figs. (\ref{mean_plot}) and (\ref{std_plot}). 

In this work, the production of CIC is assumed to be equally probable for all gain stages.  However, there may certain "hot" stages that are much more likely to produce CIC than others \cite{Bush}.  Running simulations assuming particular hot stages and comparing to this distribution will be left to future work.  

\section{Code, Data, and Materials Availability}
The script that does the analysis in this paper, as well as the data that was analyzed, is freely available at the public GitHub repository:  

{https://github.com/kjl0025/CIC\_gain\_register}

\noindent All code is in Python.

\acknowledgments 

Some of this work was done under contract with the Jet Propulsion Laboratory, California Institute of Technology.
The author would like to thank Dr. Bijan Nemati and Dr. Nathan Bush for useful discussions.

\bibliography{report} 
\bibliographystyle{spiebib} 

\section*{BIOGRAPHIES}

Kevin Ludwick is a Principal Research Scientist at the University of Alabama at Huntsville, in the Center for Applied Optics.  He does research in image processing, optics calibration, and theoretical cosmology.  He earned his Ph.D. in Physics at the University of North Carolina and was a Pirrung Postdoctoral Fellow at the University of Virginia.  He was then a professor at LaGrange College.  He has served on the executive committee for the APS FECS.

\end{document}